\documentclass[aps,pra,superscriptaddress,twocolumn,amsmath,amsfonts,amssymb,floatfix]{revtex4-1}
\usepackage{enumerate}
\usepackage{mathtools}
\usepackage{amsmath}
\usepackage{graphicx}
\usepackage{epsfig}
\usepackage{sidecap}
\usepackage{hyperref}
\usepackage{color}
\usepackage{subfigure,array}
\usepackage{verbatim}
\usepackage{multirow}

\usepackage{enumerate}
\usepackage{bm}

\begin{document}

\title{Dynamics of Bright Soliton Under Cubic-Quartic Interactions in Quasi One-Dimensional Geometry}

\author{Argha Debnath}

\author{Ayan Khan} \thanks{ayan.khan@bennett.edu.in}
\affiliation{Department of Physics, School of Engineering and Applied Sciences, Bennett University, Greater Noida, UP-201310, India}
\author{Prasanta K Panigrahi}
\affiliation{Department of Physical Sciences, IISER-Kolkata, Mohanpur, West Bengal}
\begin{abstract}
Recent inspection of liquid-like state in ultracold atomic gases due to the stabilization mechanism through the delicate balance between effective mean-field and beyond mean-field (BMF) interactions, has motivated us to study the modified/extended Gross-Pitaevskii (eGP) equation which includes the BMF contribution. In this article, we focus on variational analysis of solitonic regime with eGP equation while the soliton is subjected to an obstacle. This reveals different scattering scenarios of the soliton with explicit dependence of the BMF interaction. The results show the existence of tunneling, partial and complete trappings, in different parameter domains. These observations are further corroborated by the fast-Fourier transform method. In the later part we also extend our analysis to trapped systems. The controlled trapping in defect potential and its release can be potentially useful for quantum information storage.
\end{abstract}

\maketitle

\section{Introduction}In recent years the experimental observations of liquid-like phase in ultra-cold gases 
has led to tremendous excitement \cite{barbut2,wolfgang1,cabrera1,chomaz,bottcher,tanzi,donner}. The existence of such unique phase is attributed to the competition between
different interactions present in the system such as long range dipolar interaction \cite{barbut2}, spin-orbit coupling (SOC) \cite{wolfgang1} and short range contact 
interactions for multi component BEC \cite{cabrera1}. In most cases, the competition is between short range contact interaction with either SOC or dipolar interaction \cite{barbut2,wolfgang1,chomaz,bottcher,tanzi,donner}. However, in binary BEC, we can observe similar competition between intra and inter species interactions as well \cite{cabrera1}.

The ultra-cold atomic gases of bosonic species, forming the Bose-Einstein condensate (BEC), 
are well described via Gross-Pitaevskii (GP) equation in the realm of MF interaction \cite{dalfovo}. However, recent observations 
of the above mentioned exotic phases have led us to think beyond the well known formalism. Theoretically, it is now understood that  
the stabilization mechanism of this unique state necessitates 
beyond mean-field (BMF) contribution \cite{petrov1,petrov2}, which was introduced way back in the middle of the last century \cite{lee}. 
The new findings point towards the importance of the BMF interaction and GP equation is suitably modified to describe them. 
This has infused new interest in ultra-cold gas research \cite{yukalov,debnath4}.

Theoretically, one of the natural outcomes of GP equation is robust localized solutions, commonly known as solitons.
Solitons are exceptionally stable wave phenomena that appear in a variety
of physical systems. They propagate without spreading, due to a fine 
balance between the dispersion and the non-linearity of the physical system
\cite{weller1, middelkamp1, liu1, nistazakis1}. The property of retaining its shape over a long distance 
encourages the use of solitonic pulses in optical fibre communication.
Hence, it is evident that the propagation dynamics is very crucial for the technical application of solitons.
Therefore, it is a matter of natural curiosity to study the response of these solitary waves against impurities present in the system. 
The dynamics of bright solitons, derived from the cubic nonlinear
Schr\"odinger equation with a spatially localized (point) defect in the medium
through which it travels, have been investigated \cite{goodman1}. It has been observed that the soliton may be
reflected, transmitted, or captured by the defect in inhomogeneous condensates.
Further, the investigations of scattering of bright solitons in a
BEC due to narrow attractive potential wells and barriers have been reported \cite{thomas}.
There are observations of wave or particle-like behaviours in solitons due to the competition between 
the dispersive wave propagation and the attractive inter-atomic interaction. We also note the reporting of quantum tunnelling
of a bright soliton in ultracold atoms through a delta potential \cite{wang1}. In that process one can observe an intermediate regime where the classical and
quantum properties are combined via a finite discontinuity in the tunnelling
transmission coefficient. Soliton dynamics has a profound impact in quantum computation as well.
Recent studies suggest creation of controlled-NOT gate via soliton scattering of a potential well \cite{khawaja2} and creation of Hadamard gate 
based on superposition of composite solitons \cite{khawaja3}.

The experimental observation of bright soliton and its evolution to quantum liquid, have rejuvenated the interest in studying solitons and 
their transition to droplet-like state \cite{cabrera2,salasnich1,debnath1,debnath3} through a modified GP equation.
The BMF contribution in the modified GP equation introduces a quartic non-linearity in 
the equation of motion, along with the existing cubic non-linearity in a three-dimensional binary BEC \cite{cabrera1}.
It has been shown that it is possible to introduce suitable dimensional reduction to reduce the prescribed 3+1 dimensional system to 1+1 dimension, without losing the
essence of the cubic-quartic non-linearities. We have introduced this not so familiar nonlinear Schrodinger equation (NLSE) as cubic-quartic NLSE or CQNLSE and reported several analytical solutions 
for this system \cite{debnath2}. These solutions can be localized as well as non-localized based on the actual system. Keeping in mind the transition between droplet and bright soliton, as well as the effect of defects on the dynamical evolution it is important that we study the dynamics of the localized solutions in this newly established CQNLSE.  

The main objective of the present investigation is to study the dynamics of the localized solutions against a pinned impurity in quasi one dimensional geometry. 
More precisely, we plan to investigate their scattering properties in presence of the scatterer as an impurity. 
The formulation of the problem follows the prescription
of Ref.\cite{cabrera1}, where they had studied the mixture of two hyperfine
states of $^{39}K$ and assumed that both the components occupy the same
spatial mode \cite{debnath1}. This will allow us to write the BMF coupled nonlinear Schr\"odinger
equation as an effective one component equation. We reduce the 3+1
dimensional problem to 1+1 dimension, albeit preserving all the non-linear effects \cite{atre1, debnath1}. 
The Q1D system now consists of two nonlinear terms {\textit viz.}
the cubic term denotes the effective MF two-body interaction and the
quartic term signifies the signature of BMF or LHY contribution \cite{debnath1}. 
To study the scattering phenomena due to a localized defect, we have used a delta potential in our variational approach \cite{goodman1,thomas} and approximated it to 
a rectangular well or wall in the numerical calculation which relies on the split-step method based on fast-Fourier 
transform (FFT) \cite{debnath5}.

The paper is arranged in the following way; we briefly summarize the physical system and the genesis of CQNLSE in Sec.\ref{model} \cite{debnath1}.
In Sec.\ref{variation}, the additional potential is modelled as a short range potential to mimic white-noise 
and we apply the variational method (VM) to analyze the dynamics of the soliton in that environment. 
The result obtained via VM is corroborated through a split-step method based on FFT. These results are reported in Sec.\ref{fft}.
We model the delta function potential as a narrow rectangular well/barrier in our numerical calculation. In the later half we extend our discussion to trapped systems as well.
We draw our conclusion in Se.\ref{con}.

\section{Theoretical Model}\label{model}
Here, we consider a bosonic mixture of $^{39}K$ in a Q1D geometry \cite{debnath1}. 
The homonuclear set up consisting of two different hyper-fine states introduces two types of mean-field interactions. In this setup there exists inter-species and intra-species interactions where each species signifies one hyperfine state. We consider that the intra-species interactions are repulsive while the inter-species interaction is attractive. 
The competition between interspecies and intraspecies interaction leads to three distinct regimes, namely: miscible, immiscible and collapse \cite{ho1}. We can steer from one state to another
by tuning around with intra-species  ($a_{11}$ and $a_{22}$) and inter-species interactions  ($a_{12}$ or $a_{21}$). 
Recent investigations suggest that the liquid-like phase emerges in the phase boundary between miscible and collapse regimes, where $-\sqrt{a_{11}a_{22}}>a_{12}$ or $\delta a\propto(a_{12}+\sqrt{a_{11}a_{22}})$ leading to $\delta a<0$ \cite{ao1, chui1}. In this situation,
the BMF contribution becomes significant enough for $\delta a\approx0$, modifying the ground state and preventing the condensate from collapsing. 
From the calculation of Lee, Huang and Yang, it is understood that the BMF contribution $\delta a'\propto(\sqrt{a_{11}a_{22}})^{5/2}$ \cite{cabrera2}. 
In this region, it is acceptable to describe the system with an effective single component equation of motion provided we neglect the  
the spin excitations, resulting in the effective 
equation of motion:
\begin{eqnarray}
i\hbar\frac{\partial\Xi}{\partial t}&=&\left[\left(-\frac{\hbar^2}{2m}\nabla^2+\mathcal{V}\right)+\mathcal{U}|\Xi|^2+\mathcal{U}'|\Xi|^3\right]\Xi,\nonumber\\
&&\label{3dbgp}
\end{eqnarray}
where, $\mathcal{U}=\frac{4\pi\hbar}{m}\delta a$, $\mathcal{U}'=\frac{256\sqrt{\pi}\hbar^2\delta a'}{15m}$, $m$ being the mass of the atoms. 
We notice that Eq.(\ref{3dbgp}) contains two types of nonlinearities, the usual cubic nonlinearity as well as an additional quartic nonlinearity.
Further, in Eq.(\ref{3dbgp}), 
$\mathcal{V}=\mathcal{V}_{trap}+\mathcal{F}(x)$ describes the total potential comprising of trapping potential ($\mathcal{V}_{trap}$) and external potential ($\mathcal{F}(x)$). The trapping potential can be considered as a combination of transverse and longitudinal components of trapping, such that, $\mathcal{V}_{trap}=\mathcal{V}_T(y,z)+\mathcal{V}_L=\frac{m}{2}(\omega_{\perp}^2(y^2+z^2)+\omega_{0}^2 x^2)$). Here, $\omega_{\perp}$ and $\omega_{0}$ are the transverse and longitudinal trap frequencies respectively.  
We assume the linear defect in the form of $\mathcal{F}(x)=\frac{\hbar^2}{ma_{\perp}^2}f(x)$.
This setting may be embedded with the help of Feshbach resonance controlled by a focused laser beam. In order to reduce Eq.(\ref{3dbgp})  from $3+1$ dimension to $1+1$ dimensional Q1D case, we have made use of the following ansatz Ref.\cite{atre1},
\begin{eqnarray}
\Xi(\mathbf{r}, t) &=& \frac{1}{\sqrt{2\pi a_B}a_{\perp}}\psi\left(\frac{x}{a_{\perp}},\omega_{\perp}t\right) e^{\left(-i\omega_{\perp}t-\frac{y^2+z^2}{2a_{\perp}^2}\right)},\nonumber\\\label{ansatz1}
\end{eqnarray}
where, $a_B$ is the Bohr radius.

Applying the ansatz from Eq.(\ref{ansatz1}) in Eq.(\ref{3dbgp}) we obtain the Q1D extended GP equation as noted below,
\begin{eqnarray}\nonumber
i\frac{\partial\psi(x,t)}{\partial t}&= &\left[ - \frac{1}{2}\frac{\partial^2\psi(x,t)}{\partial x^2} + \frac{1}{2} K x^{2}+f(x)+g |\psi(x,t)|^2\right.\nonumber\\
&&\left.+g' |\psi(x,t)|^3\right]\psi(x,t)\label{bgp},
\end{eqnarray}
where, $g=2\delta a/a_B$, 
$g'=(64\sqrt{2}/15\pi)\delta a'/(a_B^{3/2}a_{\perp})$ and $K = \omega^{2}_{0}/\omega^{2}_{\perp}$ \cite{debnath3}. 
Here, it is important to note that $x$ and $t$ are now actually dimensionless, i.e., $x\equiv x/a_{\perp}$ and $t\equiv\omega_{\perp}t$. From here onward, we will follow this dimensionless notation of $x$ and $t$.

In cigar-shaped BEC, $\omega_{\perp}$ is typically more than $10$ times
stronger compared to $\omega_{0}$ \cite{khaykovich1}.
This ensures that the interaction energy of the atoms is much lower than the kinetic energy in the transverse
direction resulting $K\rightarrow0$ \cite{debnath1,debnath3}. 
In a recent study we have already demonstrated the existence of the analytical solution for Eq.(\ref{bgp}) while assuming $K=0$ and $f(x)=0$ \cite{debnath1}. 
The solution also revealed that the localized structure can 
only sustain if and only if $g<0$ or the effective mean-field interaction is attractive.
Here, we extend our analysis further to explicate the response of the solitonic wave against static impurity.
In the next section, we study the issue through a variational method where we consider a white noise like static impurity.
Later, we extend our investigation by employing FFT method, where we approximate delta function impurity with a finite size potential well/barrier. Then we corroborate both the results. 
At the later stage we also consider $K\neq0$ in our FFT calculation and briefly describe its effect.

\section{Variational Approximation}\label{variation}
The propagation dynamics of localized modes like solitons in presence of impurity is matter of intense interest from fundamental as well as technical perspectives.
Here, we intend to study the interaction dynamics of soliton with point defect while it is propagating \cite{goodman1,thomas}. In precise, we plan to solve Eq.(\ref{bgp}) (for $K=0$), while the localized white noise like defect is noted as, 
\begin{eqnarray}\nonumber
f(x)&=&V_0\delta(x)
\end{eqnarray}
Here, $V_0$ can be positive (transforming the delta potential into a wall) or negative (transforming the delta potential into a well). The interaction can lead to different interesting scenarios 
such as transmission, reflection or even trapping of the wave.

We start with an an ansatz wave function such that, 
\begin{eqnarray}
\psi&=&\psi_s+\psi_t
\end{eqnarray}
where, one mode is a free soliton ($\psi_s$) and the other mode is known as trapped ($\psi_t$).
We define them as,
\begin{eqnarray}
\psi_s&=&A_s\textrm{sech}\left(A_sx-Q_s\right)e^{i\left(V_sx+\phi_s\right)}\\
\psi_t&=&A_t\textrm{sech}\left(x/a_t\right)e^{i\left(\sigma_t\log\textrm{cosh}\left(x/a_t\right)+\phi_t\right)}
\end{eqnarray}

We solve the CQNLSE to observe the time dynamics of soliton with the help of this ansatz.
Here, the soliton mode amplitude $A_s$, position $Q_s$, velocity
$V_s$ and phase $\phi_s$ appear as time-dependent variational parameters. For the trapped mode we choose amplitude
$A_t$, healing length $a_t$ and phase $\phi_t$ as variational parameters. The trapped part contains a particular form of chirping
as $i\sigma_t\log\textrm{cosh}\left(x/a_t\right)$ \cite{sakaguchi1,thomas} which assists in the breathing mode of the soliton.

Even though the variational approach is not exact, nevertheless it provides deep insight into the mechanism involved. 
The system Lagrangian can now be expressed as,
\begin{eqnarray}\nonumber
\mathcal{L}&=&\frac{i}{2}\left(\psi^\dagger\frac{d}{dt}\psi-\psi\frac{d}{dt}\psi^\dagger\right)-\frac{1}{2}\left|\frac{\partial\psi}{\partial x}\right|^2-f(x)\left|\psi\right|^2\\\nonumber
&&-\frac{1}{2}g\left|\psi\right|^4-\frac{2}{5}g^\prime\left|\psi\right|^5\\
\end{eqnarray}
The effective Lagrangian of the system is,
\begin{eqnarray}\nonumber
L&=&\int_{-\infty}^{\infty}dx\mathcal{L}\nonumber\\
&=&-2A_s\dot{\phi_s}-2\dot{V_s}Q_s-2A_t^2a_t\dot{\phi_t}-2\left[2-\log(4)\right]A_t^2a_t\dot{\sigma_t}\nonumber\\
&&+A_t^2\sigma_t\dot{a_t}-\frac{1}{3}A_s^3-A_sV_s^2-\frac{A_t^2}{3a_t}(1+\sigma_t^2)-\frac{2}{3}gA_s^3\nonumber\\
&&-\frac{2}{3}ga_tA_t^4-\frac{6}{40}\pi g^\prime a_tA_t^5\nonumber\\
&&-\frac{3}{10}g^\prime A_s^4\left[\textrm{e}^{Q_s}\sqrt{\textrm{e}^{-2Q_s}}\textrm{arctan}\sqrt{\textrm{e}^{-2Q_s}}\right.\nonumber\\
&&\left.+\textrm{e}^{-Q_s}\sqrt{\textrm{e}^{2Q_s}}\textrm{arctan}\sqrt{\textrm{e}^{2Q_s}}\right]-V_0\left[A_t^2\right.\nonumber\\
&&\left.+2A_sA_t\textrm{sech}\left(Q_s\right)\textrm{cos}\left(\phi_s-\phi_t\right)+A_s^2\textrm{sech}^2\left(Q_s\right)\right]\nonumber\\
\end{eqnarray}

To obtain the equations of motion, we solve the following Euler-Lagrange equations \cite{goldstein},
\begin{eqnarray}\nonumber
\frac{d}{dt}\left(\frac{\partial L}{\partial \dot{q_i}}\right)&=&\frac{\partial L}{\partial q_i}
\end{eqnarray}
where, $q_i=A_s, \phi_s, Q_s, V_s, A_t, \phi_t, a_t, \sigma_t$ are the generalized coordinates \cite{thomas}. 
\begin{figure}[tbp]
\includegraphics[scale=0.35]{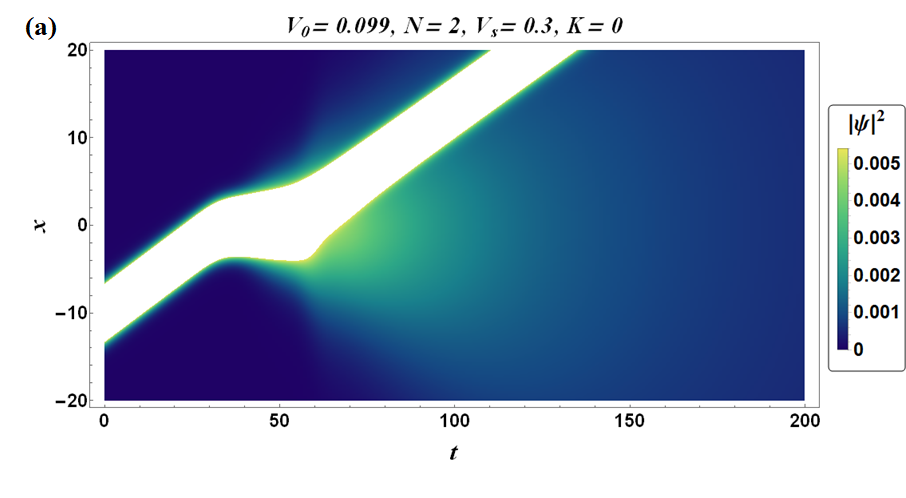}
\includegraphics[scale=0.35]{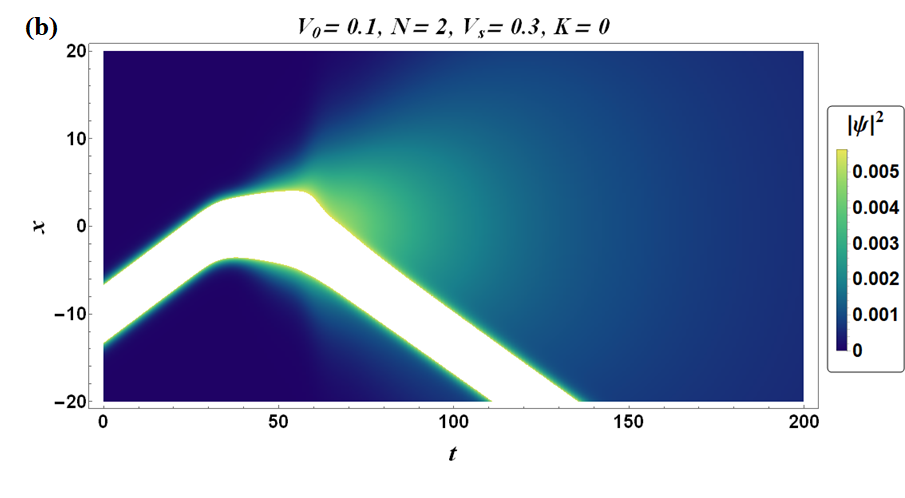}
\caption{Solitonic wave density as a function of time
for $V_0 = 0.099$, panel (a)
shows complete transmission of soliton through $\delta$-function potential wall, while panel (b) for $V_0 = 0.1$ shows the total reflection of the soliton.}\label{pvariat}
\end{figure}
The resulting dynamical equations are,
\begin{eqnarray}
\dot{A_s}&=&-V_0 A_sA_t\textrm{sech}\left(Q_s\right)\textrm{sin}\left(\phi_s-\phi_t\right),\label{ELE1}\\
\dot{\phi_s}&=&-\frac{1}{2}\left(A_s^2-V_s^2\right)-V_0[A_s\textrm{sech}^2\left(Q_s\right)+\nonumber\\
&&A_t\textrm{sech}\left(Q_s\right)\textrm{cos}\left(\phi_s-\phi_t\right)]-gA_s^2-\nonumber\\
&&\frac{6}{10}g^\prime A_s^3\left[\textrm{e}^{Q_s}\sqrt{\textrm{e}^{-2Q_s}}\textrm{arctan}\sqrt{\textrm{e}^{-2Q_s}}\right.\nonumber\\
&&\left.+\textrm{e}^{-Q_s}\sqrt{\textrm{e}^{2Q_s}}\textrm{arctan}\sqrt{\textrm{e}^{2Q_s}}\right],\label{ELE2}\\
\dot{Q_s}&=&A_sV_s,\label{ELE3}\\
\dot{V_s}&=&V_0\left[A_s^2\textrm{sech}^2\left(Q_s\right)\textrm{tanh}\left(Q_s\right)+\right.\nonumber\\
&&\left.A_sA_t\textrm{sech}\left(Q_s\right)\textrm{tanh}\left(Q_s\right)\textrm{cos}\left(\phi_s-\phi_t\right)\right],\label{ELE4}\\
\dot{a_t}&=&\frac{2\sigma_t}{3a_t}-2\left[2-\log(4)\right]\frac{V_0A_s}{A_t}\textrm{sech}\left(Q_s\right)\textrm{sin}\left(\phi_s-\phi_t\right),\nonumber\\\label{ELE5}\\
\dot{A_t}&=&\frac{V_0A_s}{2a_t}\textrm{sech}\left(Q_s\right)\textrm{sin}\left(\phi_s-\phi_t\right)-\frac{A_t\dot{a_t}}{2a_t},\label{ELE6}\\
\dot{\sigma_t}&=&\frac{1}{\left[5-2\log(4)\right]}\left[-\frac{2\sigma_t\dot{A_t}}{A_t}-2\dot{\phi_t}+\frac{1}{3a_t^2}+\frac{\sigma_t}{3a_t}-\frac{2}{3}gA_t^2\right.\nonumber\\
&&\left.-\frac{6}{10}\pi g^\prime A_t^3\right],\label{ELE7}
\end{eqnarray}
\begin{eqnarray}
\dot{\phi_t}&=&-\dot{\sigma_t}\left[2-\log(4)\right]+\frac{\dot{a_t}\sigma_t}{2a_t}-\frac{1}{6a_t^2}\left[1+\sigma_t^2\right]+\frac{2}{3}gA_t^2\nonumber\\
&&-\frac{3}{16}\pi g^\prime A_t^3+\frac{V_0}{2A_ta_t}\left[A_t+A_s\textrm{sech}\left(Q_s\right)\textrm{cos}\left(\phi_s-\phi_t\right)\right].\nonumber\\\label{ELE8}
\end{eqnarray}
We solve the eight coupled first order ordinary differential equations numerically with appropriate initial conditions \cite{wolfram}. The numerical scheme is relied upon the 
iterative method using the initial conditions. Our results agree well with Ref.~\cite{thomas} in absence of BMF interaction. Here, we like to point out that, Eqs.(\ref{ELE2}, \ref{ELE7}, \ref{ELE8}) carry explicit dependence of $g'$, while other equations have implicit dependence.

To solve the above set of equations, we set the initial velocity of free soliton to
$V_s= 0.3$ with the initial position $Q_s =-10$.
We choose small initial values for the amplitude of the trapped mode $A_t =10^{-4}$ and it's healing length 
$a_t =10^{-2}$. Furthermore, we consider the phase difference $\Delta\phi=\phi_t-\phi_s=0$ and $\sigma_t=1$ at $t=0$. 
However, We observe that the results are not affected 
by the initial choices of $\phi_s$ and $\phi_t$. 
We also set the $g=-1$ and $g^\prime=0.33$
making the cubic interaction to be attractive and quartic interaction to be repulsive in nature.

For repulsive delta potential, at $V_0=0.099$, the soliton is being completely transmitted
[Fig.~\ref{pvariat}(a)], while full reflection [Fig.~\ref{pvariat}(b)]
can be observed for $V_0\geq0.1$. 
\begin{figure}[h!]
\includegraphics[scale=0.35]{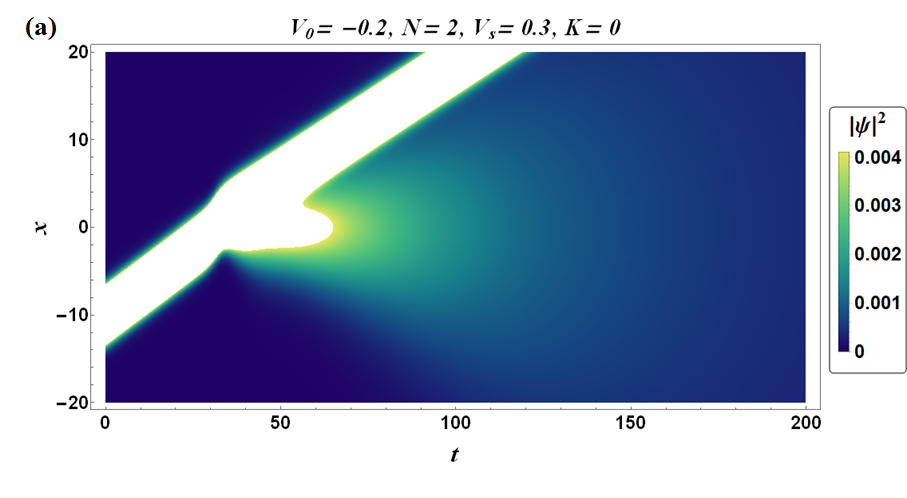}
\includegraphics[scale=0.35]{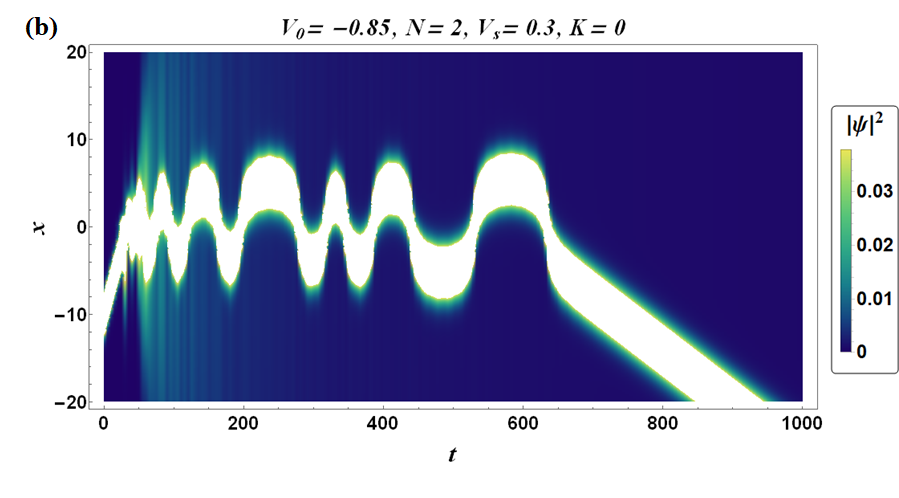}\\
\includegraphics[scale=0.35]{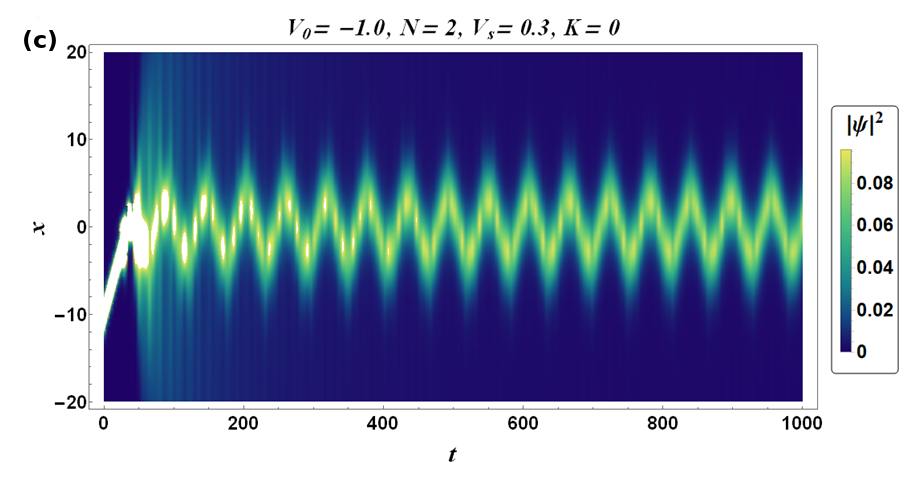}
\includegraphics[scale=0.35]{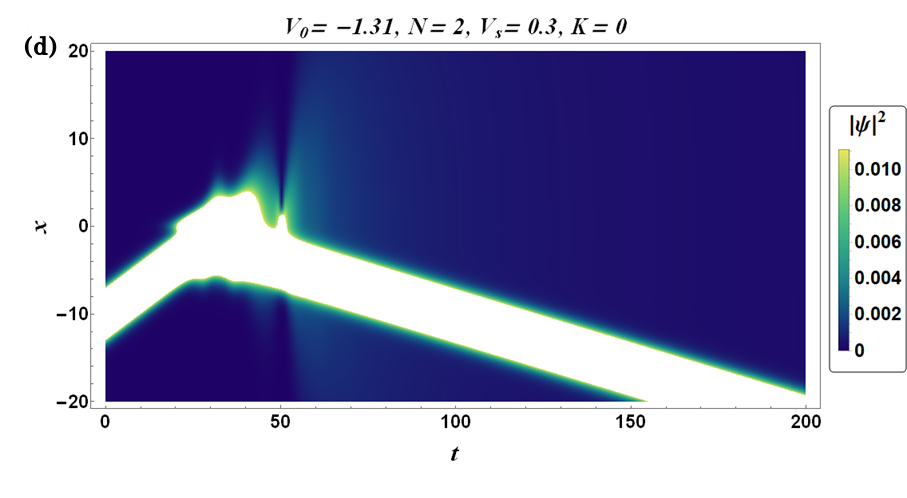}
\caption{Density plot of the solution as a function of time
by solving Eqs.(\ref{ELE1}-\ref{ELE8}). For $V_0 =-0.2$, panel (a)
described full transmission of soliton. Panel (b) depicts
partial trapping at $V_0 =-0.85$. The soliton 
is being trapped for some  period of time before it leaves the trap. 
Panel (c) shows the soliton is completely trapped at $V_0=-1.0$
and sloshes around
the well. For $V_0 =-1.31$, the whole solitonic wave is totally reflected which is depicted in panel (d).}\label{vnegative_plot}
\end{figure}
On the other hand, 
For attractive delta potential, at $V_0=-0.2$, the soliton is being transmitted
almost completely as described Fig.~\ref{vnegative_plot}(a), while full reflection can be seen in Fig.~\ref{vnegative_plot}(d) for
$V_0=-1.31$. Reflection of the soliton is seen due to the repulsive force 
induced  between two modes by the total interaction originating from MF and BMF terms when they
are out of phase. Soliton gets reflected if repulsive effect dominates over the attractive potential.
This behaviour is analogous to the situation when two coherent solitons interact with each other and
they overlap while they are in phase or avoid collision when out of phase \cite{khawaja1}.

We find a more complicated and interesting behaviour when we vary $V_0$
in between the observed reflection and transmission potential strengths.
One observes partial trapping of the soliton at $V_0=-0.85$ [Fig.~\ref{vnegative_plot}(b)] for some time
before it gets out from the trap. However, a small increment in $V_0$ can induce complete trapping as depicted in Fig.~\ref{vnegative_plot}(c). 
Progressive increase in $V_0$ results 
higher frequency of oscillation, such as for $V_0=-0.85352, -0.96, -1.0$ the frequency is $0.02, 0.04, 0.1$ respectively.
We conclude that for small $V_0$ there is full transmission, for very large $V_0$, the soliton reflects completely,
then partial and full trapping lies in between them.
\subsection*{Energy Calculation}
Here, we analyze the role of energy in trapping process in presence of attractive impurity potential. The corresponding energy functional can be written as \cite{debnath1},
\begin{eqnarray}
E[\psi(x)]&=&\int dx\left[\frac{1}{2}\left|\frac{\partial}{\partial x} \psi(x)\right|^2+f(x)|\psi(x)|^2+\frac{g}{2}|\psi(x)|^4\right.\nonumber\\
&&\left.+\frac{2g^\prime}{5}|\psi(x)|^5\right].
\end{eqnarray}\label{energyfun}
The first term describes the quantum pressure which contributes to the kinetic energy,
\begin{eqnarray}\nonumber
E_{\textrm{kin}}^\textrm{d}&=&\frac{1}{2}\int dx\left|\frac{\partial}{\partial x} \psi(x)\right|^2\nonumber\\
&=&\frac{A_s}{3}\left(A_s^2+3V_s^2\right)+\frac{A_t^2}{3a_t}\left(1+\sigma_s^2\right).
\end{eqnarray}
$E_{\textrm{kin}}^\textrm{d}\neq0$ even when the soliton is stationary.
\begin{figure}[h!]
\includegraphics[scale=0.2]{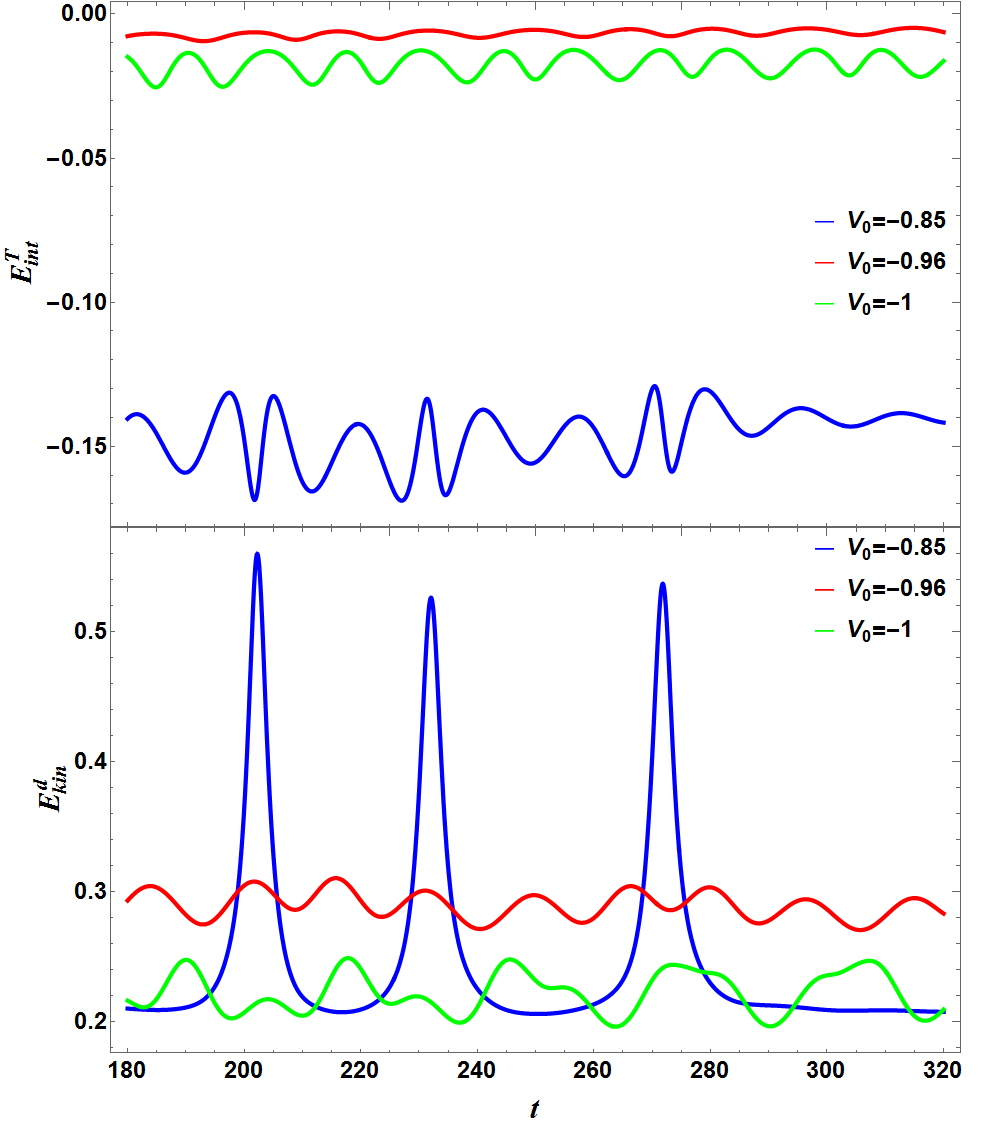}
\caption{Time dynamics of different energy contributions is depicted for the partially trapped soliton for $V_0 =-0.85$ and fully trapped soliton for $V_0 =-0.96$ and $V_0=-1$. 
The top panel describes the total interaction energy while the bottom panel represents the quantum pressure contribution in the kinetic energy.}\label{allenergy_plot}
\end{figure}

The contribution due to interactions in the energy functional can be considered as a summation of qubic and quartic interaction energy which can be noted as,  
\begin{eqnarray}\nonumber
E_{\textrm{int}}^\textrm{T}&=&E_{\textrm{int}}^\textrm{cubic}+E_{\textrm{int}}^\textrm{quartic}\nonumber\\
&=&\frac{g}{2}\int dx\left| \psi(x)\right|^4+\frac{2g^\prime}{5}\int dx\left| \psi(x)\right|^5\nonumber\\
&=&\frac{2}{3}g\left(A_s^3+a_tA_t^4\right)+\frac{6}{40}\pi g^\prime a_tA_t^5+\nonumber\\
&&\frac{3}{10}g^\prime A_s^4\left[\textrm{e}^{Q_s}\sqrt{\textrm{e}^{-2Q_s}}\textrm{arctan}\sqrt{\textrm{e}^{-2Q_s}}\right.\nonumber\\
&&\left.+\textrm{e}^{-Q_s}\sqrt{\textrm{e}^{2Q_s}}\textrm{arctan}\sqrt{\textrm{e}^{2Q_s}}\right]
\end{eqnarray}

Usually, the contribution of $E_{\textrm{int}}^\textrm{quartic}$ in total interaction energy $E_{\textrm{int}}^{\textrm{T}}$ and
ultimately in total energy $E$ is negligible as $g>>g^\prime$.
In a two-component BEC, attractive MF and repulsive BMF interactions can be
manipulated to compete with each other, with the help of Feshbach resonance.
For a given density (which is a function of $g$ and $g^\prime$) BMF energy remains
relatively large even as the MF term
shrinks. It is possible to go into regimes, where the competition of MF and
BMF energy give rise to the formation of a self-bound quantum
liquid droplet and a flat-top starts to arise, indicating constant density in bright soliton-like profile. Here we are interested in solitonic movement before the quantum phase transition in cubic-quartic interacting environment.

Time dynamics of kinetic ($E_{\textrm{kin}}^{\textrm{d}}$) and interaction
energies ($E_{\textrm{int}}^{\textrm{T}}$) are plotted in Fig.~\ref{allenergy_plot}
for different potential strengths. We are more interested in the variation of
$E_{\textrm{kin}}^{\textrm{d}}$ and  $E_{\textrm{int}}^{\textrm{T}}$ 
with time in case of partial and full trapping. For the partially trapped soliton
at  $V_0 =-0.85$, $E_{\textrm{kin}}^{\textrm{d}}$ undergoes breathing 
oscillations after scattering process while $E_{\textrm{int}}^{\textrm{T}}$ 
oscillates around a mean energy value ($\approx-0.15$). However, getting out
of the trap $E_{\textrm{kin}}^{\textrm{d}}$ stops oscillating and takes up a
constant value. On the other hand, $E_{\textrm{int}}^{\textrm{T}}$ still oscillates
after getting out of the trap but with lower amplitude and frequency.
In case of fully trapped soliton for $V_0 =-0.96$ and $V_0=-1$, $E_{\textrm{int}}^{\textrm{T}}$
oscillates taking an energy value very close to zero whereas $E_{\textrm{kin}}^{\textrm{d}}$ dominates
in the overall total energy contribution.

\section{Numerical Analysis}\label{fft}
In this section we simulate the dynamics of the soliton in a box with hard walls.
We compute Eq.(\ref{bgp}) using the split-step method based on FFT. More precisely, we have used a finite difference spatial discretizations algorithm. In computation, we have employed hard wall boundaries which allows us to use pseudospectral derivatives for finite difference calculation. The computation for pseudospectral derivative is carried out using FFT, which is efficient and minimizes round off error.

The initial wave function is considered as,
$\psi(x,t=0)= \exp(iv_sx)\psi(x-x_{int})$,
here, $x_{int}$ is initial position and $v_s$ is the initial momentum. Here, we have borrowed the analytical solution obtained for Eq.(\ref{bgp}) with $f(x)=0$ from Ref.~\cite{debnath1} as a initial wave function.

The external potential or the impurity potential is considered as a rectangular well or wall such that, 
\begin{eqnarray}
f(x)=\left\{\begin{array}{c}
V_0\,\,\,\,\textrm{for}\, L<x<L+1,\\
0\qquad\qquad\,\,\,\,\textrm{otherwise,}\end{array}\right.\label{barrier}
\end{eqnarray}
A positive $V_0$ will set up a rectangular wall as well as for well it will be a negative $V_0$. 
The convergence is closely monitored and the 
reflection from the boundaries is avoided by appropriate timing of the
simulation \cite{thomas}.

Throughout the current discussion the particle number $N$ is taken as $2$ and velocity $v_s=0.3$. In our previous investigation, where we had studied the analytical solution of CQNLSE~\cite{debnath1}, we have observed transition from solitonic to droplet like state while changing the particle number. This allows us to consider $N=2$ as a suitable choice to have solitonic wave.
The interactions $g$ and $g^\prime$ are taken as $-1$ and $0.33$ respectively, maintaining the condition $|g|=3g^\prime$.
The system size is $L = 30$ and the
potential locates at $x = x_0 = L/2 = 15$.

In this following subsections we present results from numerical solutions
of Eq.(\ref{bgp}) corresponding to a soliton approaching the potential of
Eq.(\ref{barrier}). We use the already available analytical solution from Ref.\cite{debnath1} as the initial solution and consider that the soliton is being released from initial position 
$x_{int}=5$ towards the external potential spanning from $x_{left}=15$ to $x_{right}=16$, making the potential width of $1$ unit.

Next, we calculate the reflected ($R$),
and transmitted ($T$) fraction of the
soliton such that, 
\begin{eqnarray}\nonumber
R&=&\frac{1}{N}\int_{-\infty}^{L/2}{dx|\psi(x)|^2},\nonumber\\
T&=&\frac{1}{N}\int_{L/2+1}^{\infty}{dx|\psi(x)|^2},\nonumber
\end{eqnarray}
It is natural to assume that the fractions should follow the constrained condition as $R+T=1$, which we have verified at the very onset.
\subsubsection*{Effect of Defect Wall}
\begin{figure}[h!]
\includegraphics[scale=0.35]{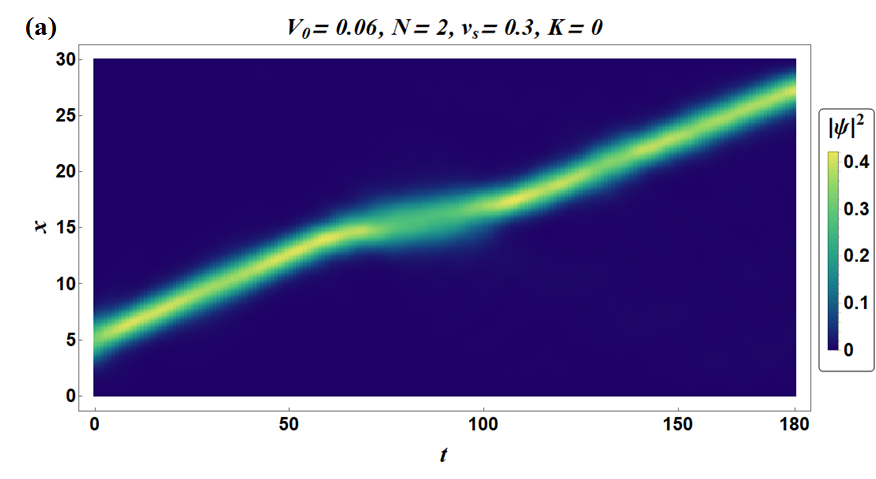}
\includegraphics[scale=0.35]{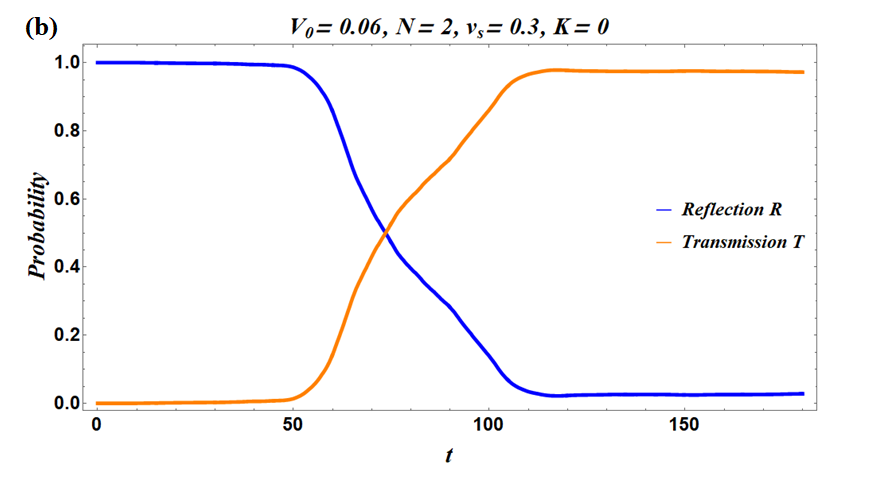}
\includegraphics[scale=0.35]{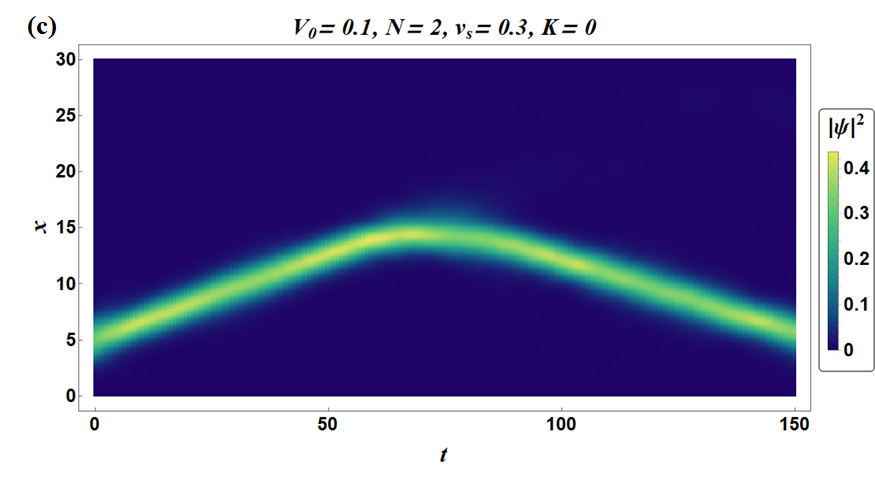}
\includegraphics[scale=0.35]{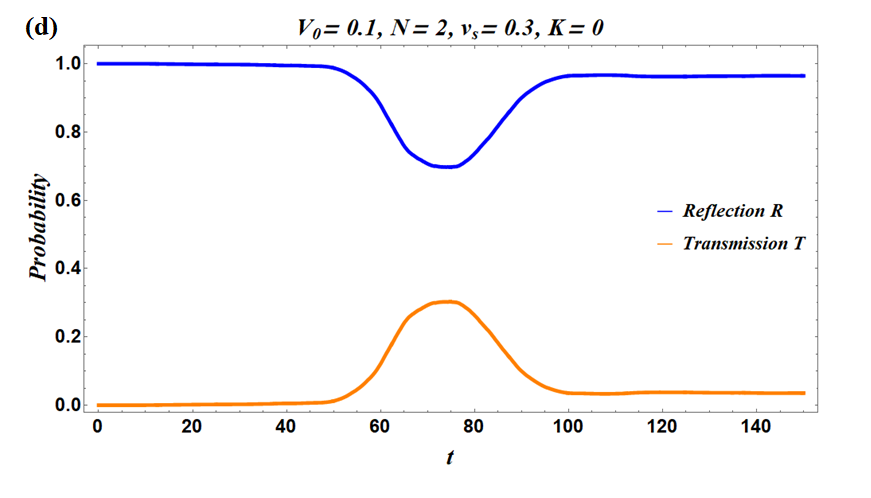}
\caption{The figures depict transmission of solitonic wave with time in presence of a wall.
Panel (a) presents total transmission at $V_0=0.06$ whereas panel (b) shows the reflected ($R$) and transmitted ($T$) fractions. With the increase of potential strength
at $V_0=0.1$, shown in panel (c), total reflection can be seen. Panel (d) complements (c) with the calculation of $R$ and $T$.}\label{fft_pos_pot}
\end{figure}
As mentioned earlier, we carry out a systematic study of the barrier strength starting from a very weak one. When we use a weak barrier or wall of reasonably smaller height say, $V_0=0.06$ we observe the
complete transmittance of the localized mode which can be noted from Fig.~\ref{fft_pos_pot}(a). The corresponding reflection coefficient and transmission coefficients are depicted 
in Fig.~\ref{fft_pos_pot}(b).     
However, the situation reverses when we use a higher barrier potential such as $V_0=0.1$ as noted in Fig.~\ref{fft_pos_pot}(c) where we observe full reflection which is supported by the calculation 
in Fig.~\ref{fft_pos_pot}(d). The figures further suggests a small probability of penetration of the modes through the left wall which eventually gets reflected from the right wall. However, gradual increase in the barrier height results in sharper reflection implying full reflection from $x_{left}$ itself. 

\subsubsection*{Effect of  Defect Well}
\begin{figure}[h!]
\includegraphics[scale=0.35]{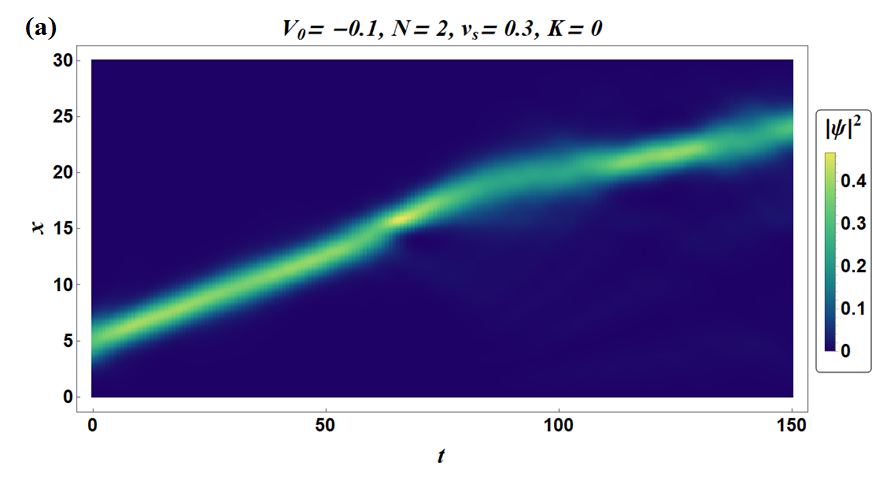}
\includegraphics[scale=0.35]{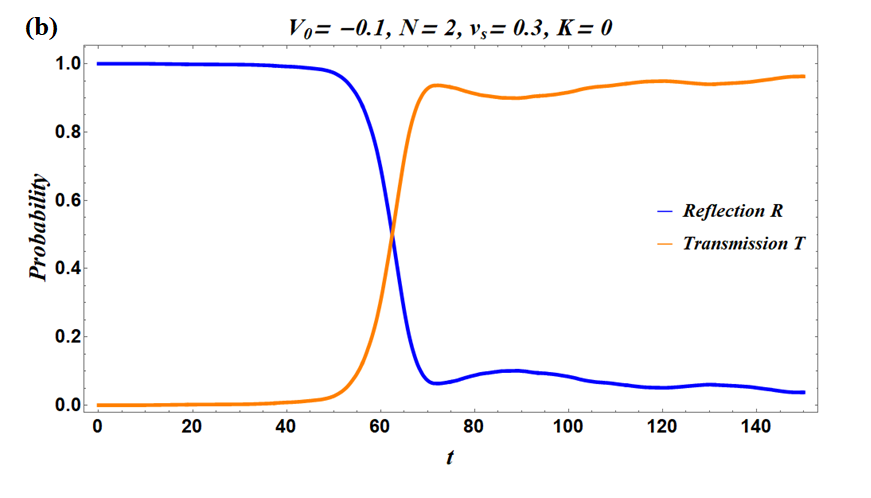}
\includegraphics[scale=0.35]{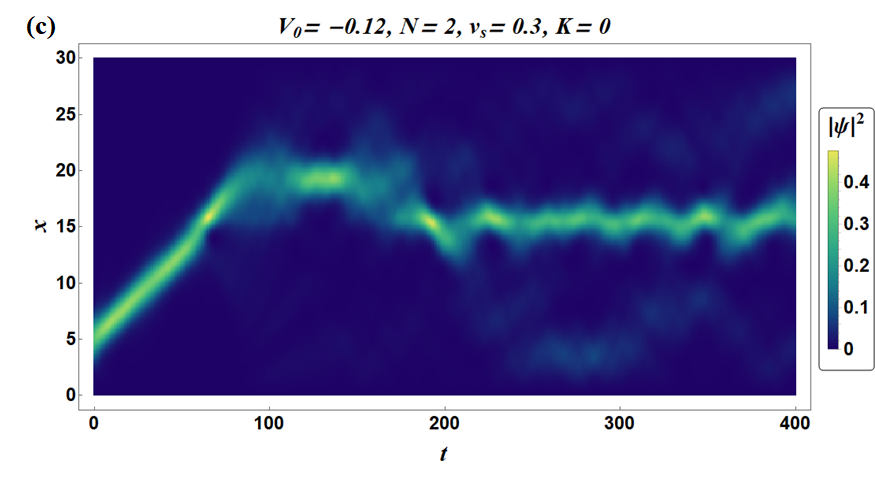}
\includegraphics[scale=0.35]{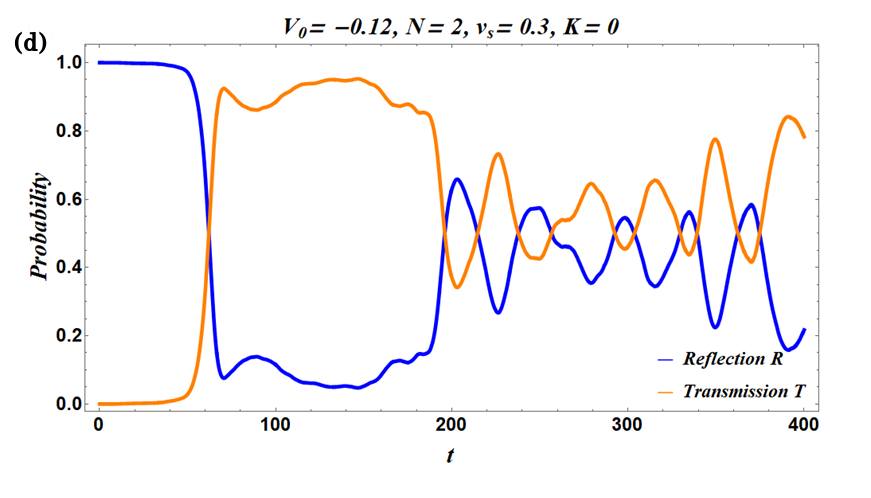}
\caption{The figures demonstrates the dynamics of localized wave in presence of a well.
 Panel (a) shows the propagation of the density wave at $V_0=-0.1$. The result is then corroborated with the calculation of $R$ and $T$ depicted in (b).
Panel (c) describes the propagation density at $V_0=-0.12$, and corresponding calculation of the reflectance and transmittance is presented in (d).}\label{fft_neg_pot_a}
\end{figure}
\begin{figure}[h!]
\includegraphics[scale=0.35]{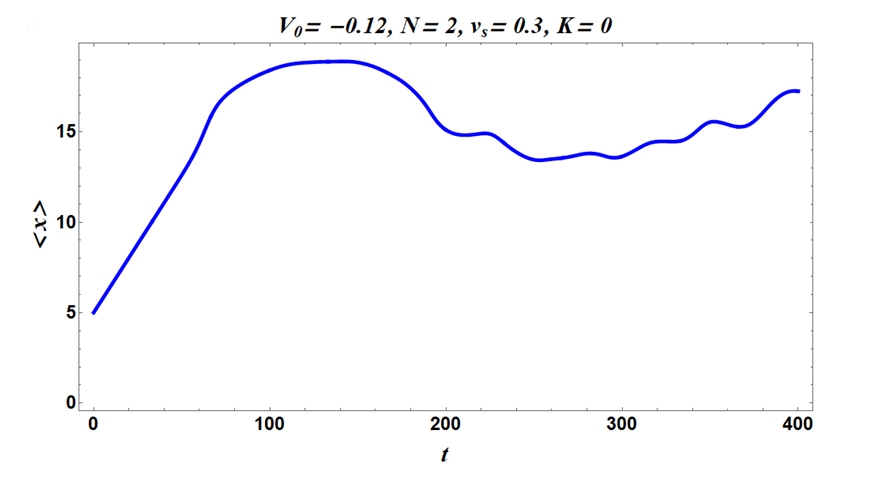}
\caption{
The time evolution of the position expectation value of ($\langle x\rangle$) of the solitonic wave for $V_0=-0.12$.}\label{fft_neg_pot_expect}
\end{figure}

In presence of a well with $V_0=-0.1$, we observe transmission of the wave as described in Fig.~\ref{fft_neg_pot_a}(a), however, in comparison with the barrier potential [Fig.~\ref{fft_pos_pot}(a)], 
we clearly see more dispersion. This is more evident when we compare the transmittance and reflectance between [as depicted in \ref{fft_pos_pot}(b) and \ref{fft_neg_pot_a}(b)].
However, $V_0=-0.12$, the wave appears more distributed as noted in Fig.~\ref{fft_neg_pot_a}(c). When we study the reflection and transmission coefficients, we observe the fluctuation in $R$ and $T$ which in order leads to trapping of the wave (partially). To investigate the trapped mode we calculate the time evolution of the position expectation value which clearly indicates that the wave is confined in a band $\sim 15\pm 4$ as described in Fig.~\ref{fft_neg_pot_expect}. Nevertheless, the the potential is placed in the region $x\in[15,16]$. Hence, we can visualize the phenomena of partial trapping as well as tunnelling of the wave.

\begin{figure}[h!]
\includegraphics[scale=0.35]{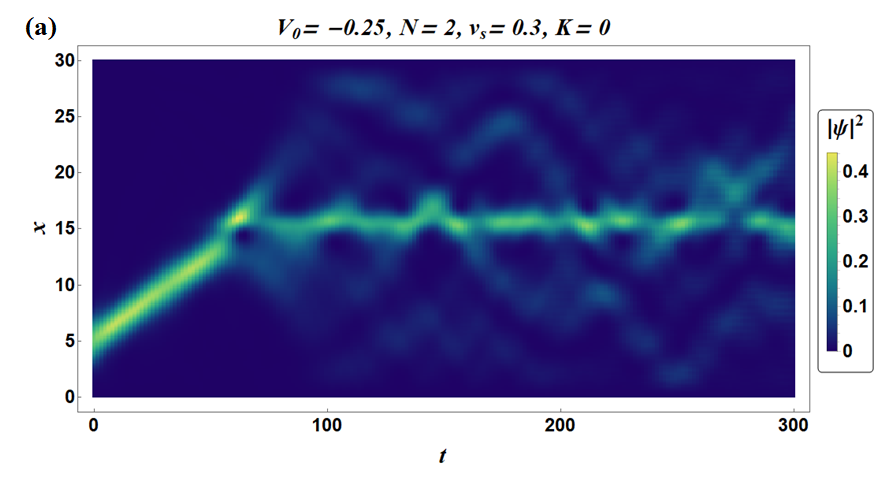}
\includegraphics[scale=0.35]{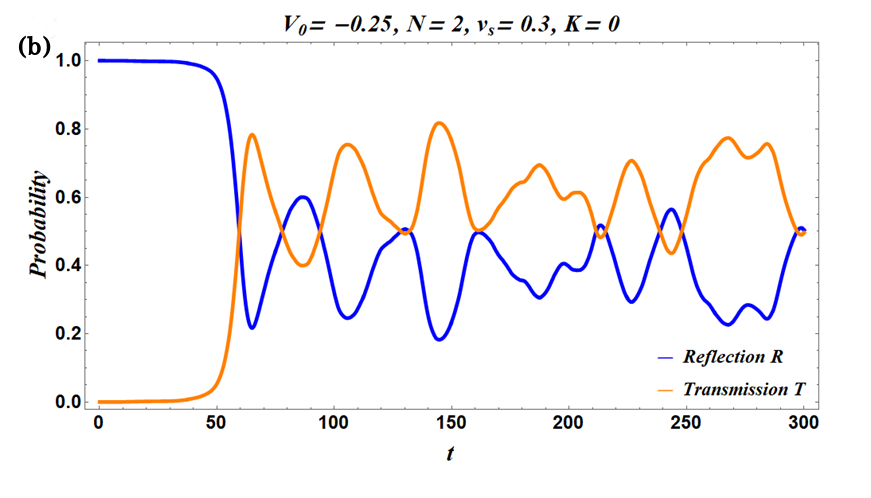}
\caption{The figures depict evolution of the wave density with time in presence of a well.
 Soliton gets trapped inside the well at $V_0=-0.25$ is shown in panel (a) and (b).}\label{fft_neg_pot_b}
\end{figure}

\begin{figure}[h!]
\includegraphics[scale=0.35]{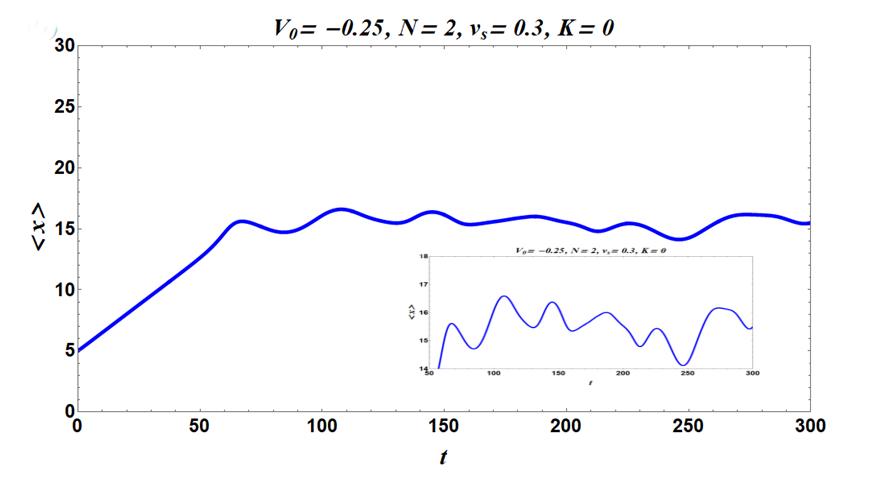}
\caption{The position expectation value of the solitonic wave is depicted which suggests the wave gets largely confined inside the potential. The inset zooms the time evaluation in the spatial dimension confirming that the wave is captured in between $x\in[15,16]$, (defined width of the potential) with very little sloshing.}\label{fft_neg_pot_osci}
\end{figure}

To observe complete trapping it is required to increase the potential further as we note that at $V_0=-0.25$ the wave appears completely trapped as depicted in Fig.~\ref{fft_neg_pot_b}(a). The corresponding time variation of $R$ and $T$ is presented in Fig.~\ref{fft_neg_pot_b}(b). However, we obtain the conclusive evidence from our calculation of $\langle x\rangle$ which is shown in Fig.~\ref{fft_neg_pot_osci}. If we focus in the inset, we can clearly see that the position expectation value is mostly confined between $x\in[15,16]$ with very little sloshing outside. 

\begin{figure}[h!]
\includegraphics[scale=0.35]{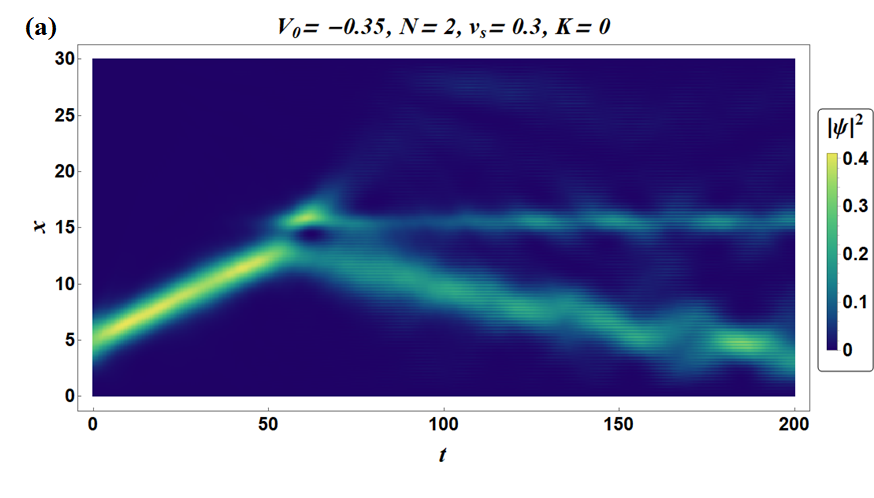}
\includegraphics[scale=0.35]{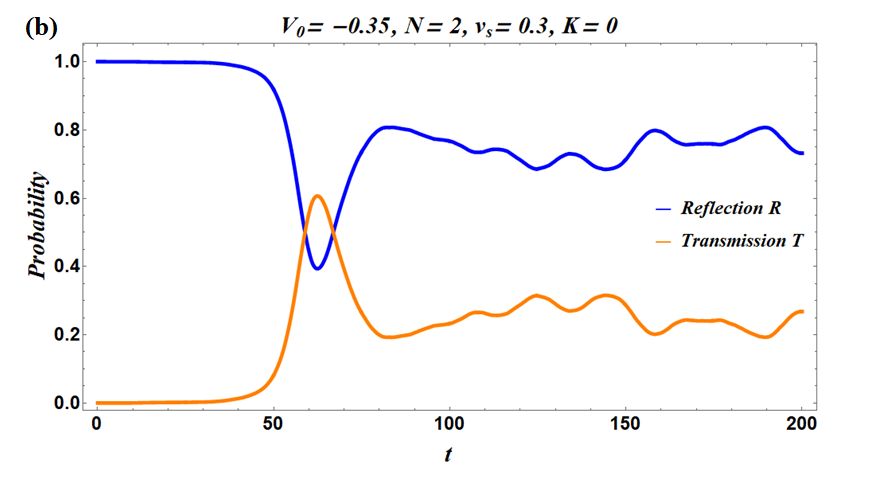}
\includegraphics[scale=0.35]{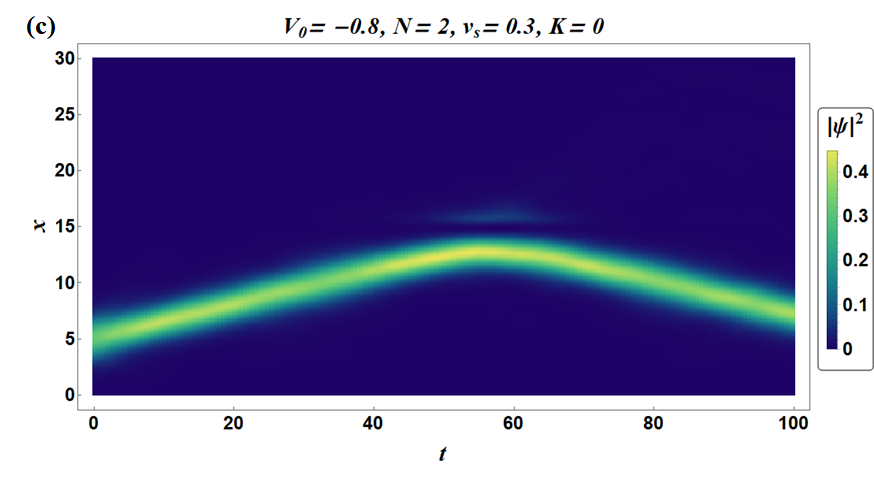}
\includegraphics[scale=0.35]{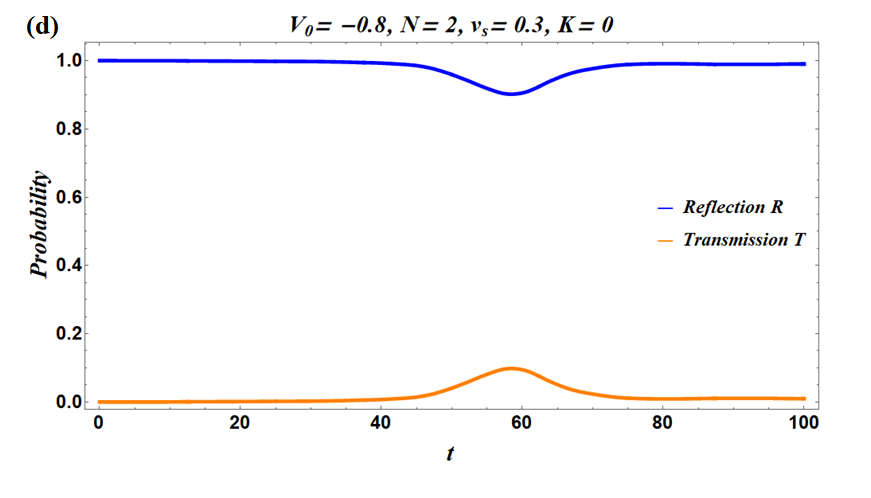}
\caption{The figures depict evolution of codensate density with time in presence of a well. At $V_0=-0.35$ partial
 trapping is observed in panel (a) and (b); While total reflection can be seen at $V_0=-0.8$ in panel (c) and (d).}\label{fft_neg_pot_c}
\end{figure}

Partial reflection
can be seen at $V_0=-0.35$  [Fig.~\ref{fft_neg_pot_c}(a)], leaving certain fraction of soliton
density inside the well to be trapped and other remaining fraction to be reflected, nonetheless we are unable to capture this feature in our VM calculation which might be a indication of the limitation of the VM in this unique system. With further increase
in the depth of the well, at $V_0=-0.8$  [Fig.~\ref{fft_neg_pot_c}(c)] we observe complete reflection. 

We have summarized and tabulated all these observations in Table \ref{t1}. From the table, we can safely conclude that 
the VM and FFT methods support each other quite well. The little discrepancy can come from the fact that the VM calculation is with a delta function potential while 
FFT calculation is based on a finite height potential. 

\begin{table}
\begin{tabular}{ | m{5em} | m{6em}| m{7em} | m{5em}| } 
\hline
Defect as a form of a & Oservable & $V_0$ in VM Method & $V_0$ in FFT Method \\ 
\hline\hline
\multirow{2}{5em}{Wall} & Transmission & $0.099$ & $0.06$ \\ \cline{2-4}
& Reflection & $0.1$ & $0.1$\\
\hline\hline
\multirow{5}{5em}{Well} & Transmission & $-0.2$& $-0.1$ \\ \cline{2-4}
 & Partial Trapping & $-0.85$& $-0.12$\\\cline{2-4}
 & Complete Trapping & $-0.86$ to $-1$ & $-0.25$ \\ \cline{2-4}
 & Complete Reflection & $-1.31$& $-0.8$\\
\hline
\end{tabular}
\caption{Comparison of $V_0$ obtained from variational analysis and fast-Fourier transform method.\label{t1}}
\end{table}
To substantiate our numerical strategies, we explicate an analytical method where we can estimate a threshold velocity which separates the transmitting and reflecting solitons due to an potential well. We have already observed from the above discussions that, by changing the strength of the potential, a smooth transition from 
transmission to trapping to reflection can be observed. Hence, we can employ an adiabatic approximation which neglects the deformation of the solitons due to a potential barrier.

Let us consider $V_0$ as a attractive delta function
potential and treating it perturbatively makes it possible to predict the threshold value $v_{thr}$ of velocity $v_s$ which
separates the rebound and passage of the incoming soliton. In the adiabatic approximation, an effective potential of the interaction of the soliton with the barrier can be expressed as \cite{debnath5},
\begin{eqnarray}
U_{\mathrm{eff}}&=&\varepsilon \int_{-\infty }^{+\infty }\delta
(x)\left\vert \psi \left( x-\zeta \right) \right\vert ^{2}dx
\end{eqnarray}%
where $\zeta $ is the instantaneous coordinate of the solitons's centre. Using the stable analytical solution from Ref.\cite{debnath1}, we can determine $U_{eff}$ and subsequently the maximum of the effective potential ($U_{\max}$) such that,
\begin{equation}
U_{\max }=\varepsilon\left(\frac{1+\mu_I }{ 1+\sqrt{\mu_I} }\right) ^{2},  \label{Umax}
\end{equation}%
where, $\mu_I=12\mu/g$. Now if we compare this expression with kinetic energy ($E_{kin}=Nv_s^2/2$), it is possible to predict the
threshold value of the velocity as,%
\begin{equation}
v_{\mathrm{thr}}^{2}=
\frac{2\varepsilon}{N}\left(\frac{1+\mu_I }{ 1+\sqrt{\mu_I} }\right) ^{2}.  \label{thr}
\end{equation}%
 The incident soliton with $v_s<v_{%
\mathrm{thr}}$ or $v_s>v_{\mathrm{thr}}$ is expected to bounce
back or pass the barrier, respectively.

It is now trivial to corroborate the analytical conclusions which our numerical results. For example, when $V_0=-0.1$, $v_{\mathrm{thr}}$ is $0.256$ which is lower than $v_s$ ($=0.3$), thus one can observe the transmission of the wave as described in Fig.~\ref{fft_neg_pot_a}(a).
Similarly, when $V_0=-0.8$, $v_{\mathrm{thr}}$ is $0.72$ which is much higher than the soliton velocity ($v_s=0.3)$. As expected, we note the reflection of the wave in Fig.~\ref{fft_neg_pot_c}(c).

\subsection*{Effect of Harmonic Trapping}
In our previous discussions, we have considered $K=0$ implying a quasi-homogeneous condition, albeit for all practical purposes $K\neq0$.
Here, we plan to highlight the effect of harmonic trapping using our numerical scheme. We have already seen that our variational approach 
fits quite well with the FFT calculation, therefore for brevity and analytical simplicity we restrict ourselves to numerical calculation only.
Nevertheless, it is possible to extend the variational approach for trapped systems as well.
\begin{figure}[h!]
\includegraphics[scale=0.35]{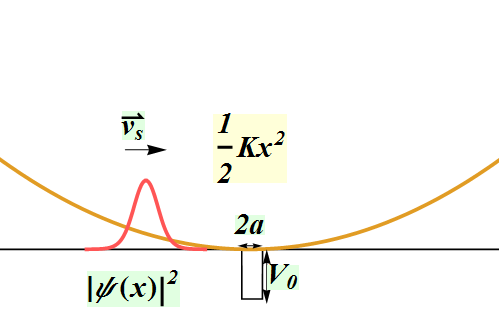}
\caption{Schematic description of the problem. The soliton is moving with a finite velocity in trap while the potential well is superposed in the same spatial dimension.}\label{cartoon2}
\end{figure}

In one of our earlier studies, we have pointed out that the competition between MF and BMF interactions results the generation of solitonic mode at low particle number and droplet formation starts with moderate particle number. At high particle density solitoninc mode again emerges \cite{debnath1}. However, in another investigation, we have also pointed out that the trapping frequency plays an important role in the nature of the solution and induces droplet to soliton transition for reasonably moderate particle number \cite{debnath3}. Here, with low particle number, which in general supports the solitonic mode, we investigate the role of the harmonic confinement. 

\begin{figure}[h!]
\includegraphics[scale=0.35]{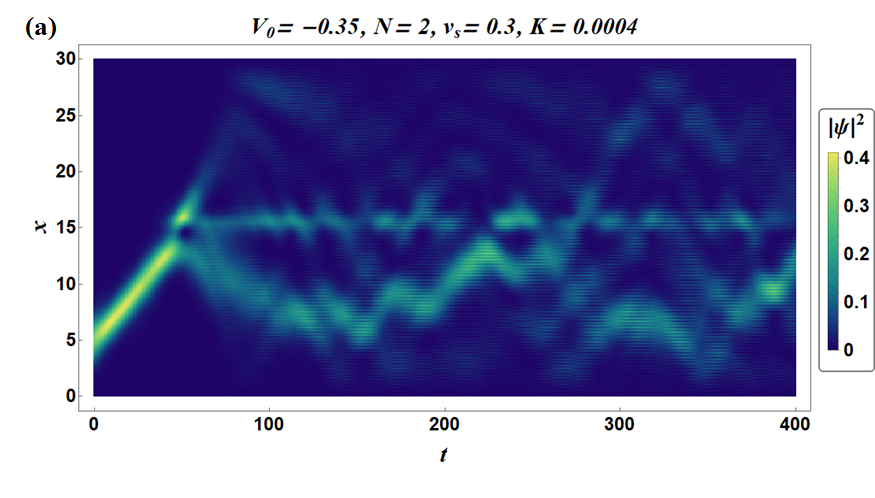}
\includegraphics[scale=0.35]{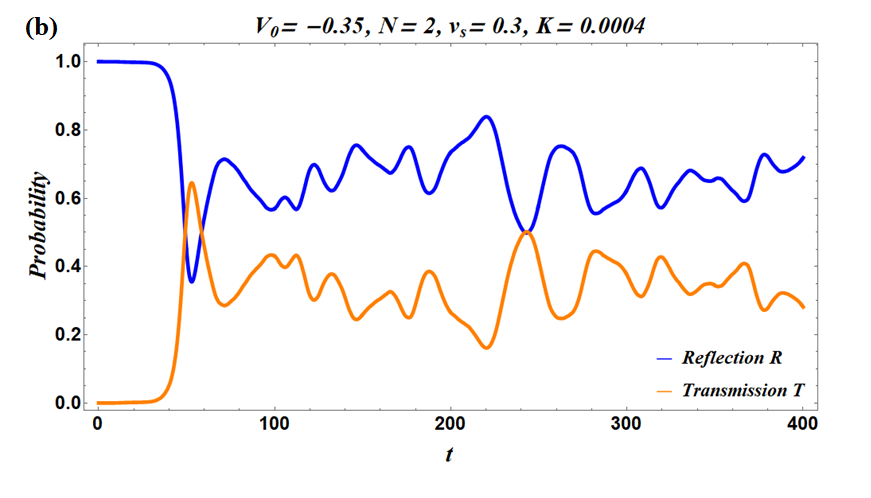}
\caption{The recombination of the splitted parts can be seen for $K=0.0004$ at $V_0=-0.35$ in panel (a); At $t\sim 240$ the recombination is observed in panel (b).}\label{fft_trap_pot}
\end{figure}

Thus, we carry out a systematic analysis for different trapping frequencies as described in Fig.~\ref{cartoon2}. In the previous subsection,  
we have seen partial reflection at $V_0=-0.35$  [Fig.~\ref{fft_neg_pot_c}(a)], splitting
the soliton in to two parts. With $K=0.0004$ [Fig.~\ref{fft_trap_pot}(a)] a recombination of the splitted
parts can be seen. The signature of recombination is also captured in Fig.~\ref{fft_trap_pot}(b) at $t\approx240$. A similar feature is also seen in Fig.~\ref{fft_neg_pot_b}
where $V_0=-0.25$. Hence, it can be concluded that the presence of weak harmonic trapping, reduces the effective potential, while competing with the attractive quantum well.
 
\begin{figure}[h!]
\includegraphics[scale=0.25]{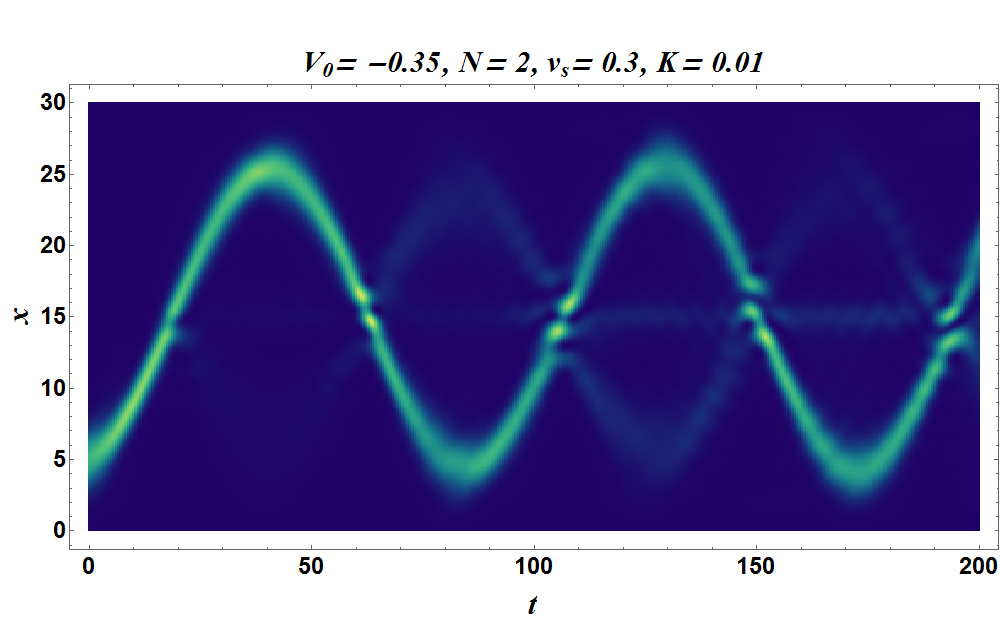}
\caption{Soliton oscillation inside the trap with minimal effect of the well at $x=15$. Here, $K=0.01$ and other parameters remain unchanged. This shows that the soliton oscillation frequency $\omega_s\sim 0.07$ which is $1/\sqrt{2}$ times the longitudinal frequency as $K=\frac{\omega_0^2}{\omega_{\perp}^2}$, and we set $\omega_{\perp}=1$. All parameters are in arbitrary units.}\label{sol_osci}
\end{figure}

On the contrary, strong trapping potential overwhelms the well and we can get back the usual soliton oscillation in the trap as noted in Fig.~\ref{sol_osci}, with soliton the oscillation frequency following the same trend as $1/\sqrt{2}$ times the longitudinal frequency \cite{fritsch}. We check our calculation for potential barrier as well, and note that for very weak longitudinal trapping and weak barrier, the solitonic mode is almost completely transmitted and increasing the barrier height induces reflection as observed for quasi homogeneous system (Fig. \ref{pvariat} [analytical method] and Fig. \ref{fft_pos_pot} [numerical method]). For strong confinement, the barrier strength becomes nearly redundant and we observe behaviour of the solitonic solution very similar to Fig. \ref{sol_osci}.

\section{Conclusion}\label{con}
In conclusion, we have studied the dynamics of soliton in Q1D geometry in presence of the BMF interaction to find how 
the travelling soliton responds to pinned obstacles on its way.
Our study throws light on the important role played by the BMF interaction on soliton dynamics in presence of defects.
In our earlier investigations, we have already pointed out the role of particle number and trap (see Ref.\cite{debnath1} and \cite{debnath3}) in determining the domains of droplet formation and soliton generation. Here, we have concentrated in the region where soliton formation is supported in Q1D. One of the main reasons to concentrate on solitonic mode in Q1D is due to the fact that this localized mode is highly sought after in information transfer. 

We begin with a three-dimensional coupled GPE along with BMF contribution. The equation is effectively single component as we assume
both the components occupy the same spatial mode in Q1D geometry \cite{cabrera2,debnath1}. The resulting Q1D equation is unique as it contains an
additional quartic nonlinearity along with the usual cubic one. The investigation is focused on the 
scattering of a bright soliton against a linear defect with the variation of its strength
via. variational approximation and  fast Fourier transform.

In VM, we observe distinct contributions of the BMF in the coupled dynamical equations for solitonic and trapped modes. We observe reflection, transmission, partial trapping and complete trapping by modulating the strength of defect $V_0$. It is to be noted that for VM calculation we have neglected the longitudinal harmonic potential. The variational calculation is corroborated numerically using the FFT method.

However, by modulating $K$, recombination of initially splited soliton can be seen.
We have summarized the observations in Table.\ref{t1}, where we observe some variation of scattering for different defect strengths for two different methods. This can be attributed to the fact that different bright soliton profiles are used as a 
solution of Eq.(\ref{bgp}) to simulate. Another difference is that in the FFT method, we have used a defect potential of finite width instead of delta function.
Nevertheless, we find that these variational results are qualitatively consistent with the numerical results obtained using the split-step FFT method.
From the FFT method we also obtain the signature of partial reflection, which the VM was unable to capture. Further, we have used FFT to study the role of the external trap potential. For weak trapping, the results are similar to that of the homogeneous system, however for strong trapping potential, the effect of pinned impurity is negligible and the solitons starts to oscillate in side the trap with $\omega_s=\omega_0/\sqrt{2}$.

We observe that the scattering properties of the soliton on a localized potential resembles that of classical
or quantum mechanical scattering, depending on the strength of the defect. When
the strength is too low it transmits, on the other hand reflects against stronger defects.
Partial and fully trapped modes are also observed for intermediate strength. Oscillation in energy can
be observed when soliton is inside the trap, indicating breathing oscillations after the
scattering process. 

These observations may be of interest from the perspective of `atom interferometry' and `quantum gates' and provide useful insight 
towards the technological domain in the years to come. 

\section*{Acknowledgement} AK thanks Department of Science and Technology (DST), India
for the support provided through the project number CRG/2019/000108. AD acknowledges the warm hospitality of IISER-Kolkata as a visitor where part of the work was carried out.
AK and AD also acknowledge insightful comments and suggestions of B. A. Malomed.

\section*{Data Availability}
The data used to support the findings of this study are included within the article.

\bibliographystyle{apsrev4-1}
\bibliography{ms_v4}

\end{document}